# Getting Beyond 5G: A Feasibility Study for 6G Using On-Chip Wireless THz Systems

**RAJAT PUSHKARNA,**[1*]

[1]*School of Electrical and Electronic Engineering, Nanyang Technological University, 50 Nanyang Avenue, 639798, Singapore*

*[*rajat002@e.ntu.edu.sg](mailto:rajat002@e.ntu.edu.sg)*

**Abstract:** The article describes a wireless system operating at 300GHz with a channel bandwidth of 40GHz. An off-chip system for wireless data transmission has been proposed and channel modelling has been done which shows the Shannon's channel capacity at 300GHz. For channel modelling both Line-of -Sight(LOS) and Non-Line of Sight(NLOS) components have been taken into account.

## 1. Introduction

Last few years have seen a significant work towards harnessing the bandwidth capabilities at lower frequencies. Many algorithms and techniques have been developed to utilize the limited bandwidth at lower frequency. As more and more connected devices are coming up the data generated is expanding exponentially which clearly shows the need of higher bandwidth to accumulate all the generated data.

Recently with the expansion of 5G using mmWave paved way to explore higher frequency asking researchers to harness much higher bandwidth at THz frequencies. The work proposed here describes an on-chip trans-receiver operating at 300GHz. The on-chip source used is a Resonance Tunneling Diode(RTD) which is the THz generator and improvising on the existing Photonic crystal on a silicon chip[1], a Photonic Topological Insulator is used[2].

## 2. On-Chip THz system for Wireless Communication

### 2.1 RTD with Photonic Topological Insulator

A Resonance Tunneling Diode works on an input power of about 1mw and have a receiver sensitivity in the range of 10μW. The other systems working in the same range uses an off-chip source to provide more input power which results in higher range of transmission [3] with input sources as UTC-PD or off chip lasers. The proposed model with PTI is a 1mW RTD operating at 300GHz and can transmit to a range of 1m.

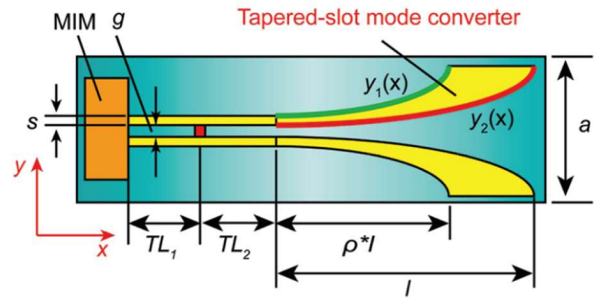

Fig. 1. A schematic model of on-chip transreceiver using PTI

Terahertz waveguiding in the conventional approaches are dependent on sharp bending and defects in the photonic crystal. Here, an all silicon chip is proposed for demonstrating the robust valley transport of light through sharp bends on a silicon based chip. The valley kink states are excellent information carriers owing to their robustness, single-mode propagation and linear dispersion [4].

*2.2 Channel modelling and wireless feasibility of Terahertz*

For any wireless system to work efficiently, channel modelling and the performance of baseband signal in the environment plays a key role. It is being demonstrated in various articles that the frequency range from 100-500GHz gets dissipated by water molecules and gets scattered by the reflective surfaces as demonstrated by the HiTRAN dataset. A study of molecular absorption has been done for the frequency range of 100-500GHz and windows are determined based on the least absorptive frequencies.

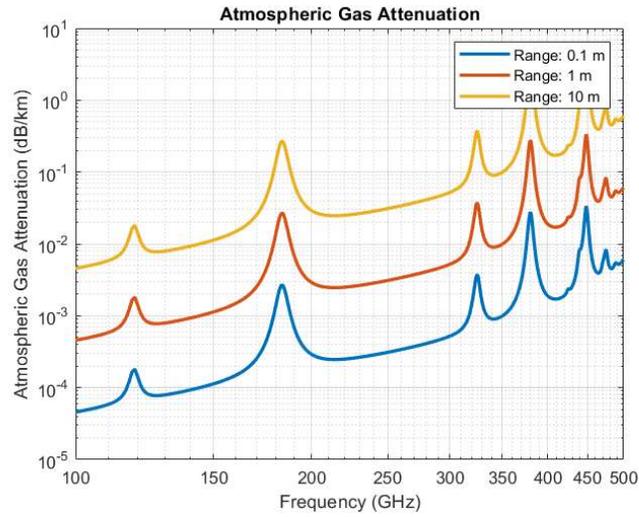

Fig. 2. Terahertz attenuation spectra

Based on the attenuation topology shown in Fig 1 a frequency window of 300GHz is selected as the baseband signal. Using 300GHz as carrier frequency, the free space path loss is computed using equation which is given as:

$$20log_{10}(d) + 20log_{10}(f) + 20log_{10}\left(\frac{4\pi}{c}\right) \qquad (1)$$

Based on the equation we can calculate the overall transmission loss from the transmitter to the receiver. Equation 1 clearly states that the path loss is dependent on two parameters, distance of transmission and the operating frequency. This can be seen clearly in the Fig 3.

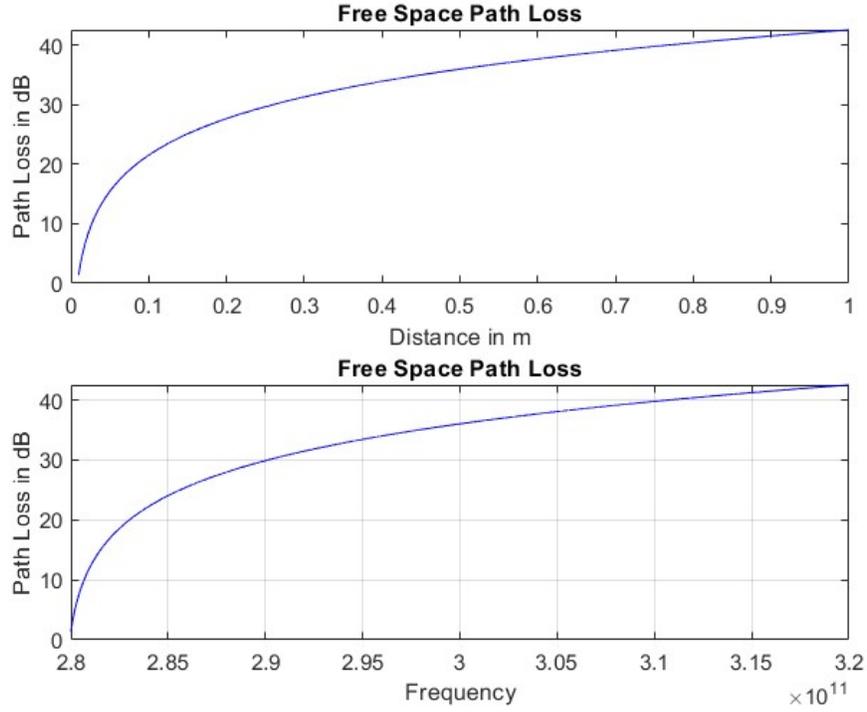

Fig. 3. Path loss model with varying the distance between receiver and transmitter and as a dependence of frequency

Once the path loss is determined at a given frequency which is 300GHz for the study. The total power which can be received at the receiver is computed. The received power can be calculated using the Friis transmission equation given as:

$$P_{rx} = P_{tx} G_{tx} G_{rx} \left(\frac{c}{4\pi D_r f}\right)^{\wedge} 0.5 \qquad (2)$$

**Table 1. Parameters for calculating theoretical received power**

| Parameters | Values |
|---|---|
| $P_{tx}$ | 0dBm |
| $G_{tx}$ | 20dBi |
| $G_{rx}$ | 20dBi |
| $D_r$ | 100cm |
| $f$ | 300GHz |

The channel modelling can be done for two scenarios one is the Line-of-Sight (LOS) and the Non-line-of-sight (NLOS) for the baseband signal. A deterministic channel model for the 0.1 to 1 THz frequency range can be modelled using Kirchhoff's scattering theory and ray tracing theory for LOS and NLOS propagation.

The magnitude of the LOS path $H^{LOS}$ (f , r) is given as:

$$H^{LOS}(f,r) = H_{spread(f,r)} \cdot H_{abs(f,r)} \quad (3)$$

where,
$$H_{spread(f,r)} = \frac{c}{4\pi \cdot f \cdot r} \quad (4)$$

and,
$$H_{abs(f,r)} = e^{-\frac{1}{2}\alpha_{molec}(f,T_k,p)r} \quad (5)$$

where, $\alpha_{molec}$ is the absorption coefficient of air, f is the operating frequency, $T_k$ is the temperature and r is the distance of transmission.

The $i^{th}$ NLOS channel model is given as

$$H_i^{NLOS}(f,r,\xi_i) = H_{refl,i}(f, r_{i2}, \theta_{i1}, \theta_{i2}, \theta_{i3}) \times H_{spread,i}(f, r_{i1}, r_{i2}) \cdot H_{abs,i}(f, r_{i1}, r_{i2}) \quad (6)$$

where,
$$H_{refl,i}(f, r_{i2}, \theta_{i1}, \theta_{i2}, \theta_{i3}) = \sqrt{E\{R_{power,i}(f, r_{i2}, \theta_{i1}, \theta_{i2}, \theta_{i3})\}} \quad (7)$$

$$H_{spread,i}(f, r_{i1}, r_{i2}) = \frac{c}{4\pi \cdot f \cdot (r_{i1}+ r_{i2})} \quad (8)$$

$$H_{abs,i}(f, r_{i1}, r_{i2}) = e^{-\frac{1}{2}\alpha_{molec}(f,T_k,p)(r_{i1}+ r_{i2})} \quad (9)$$

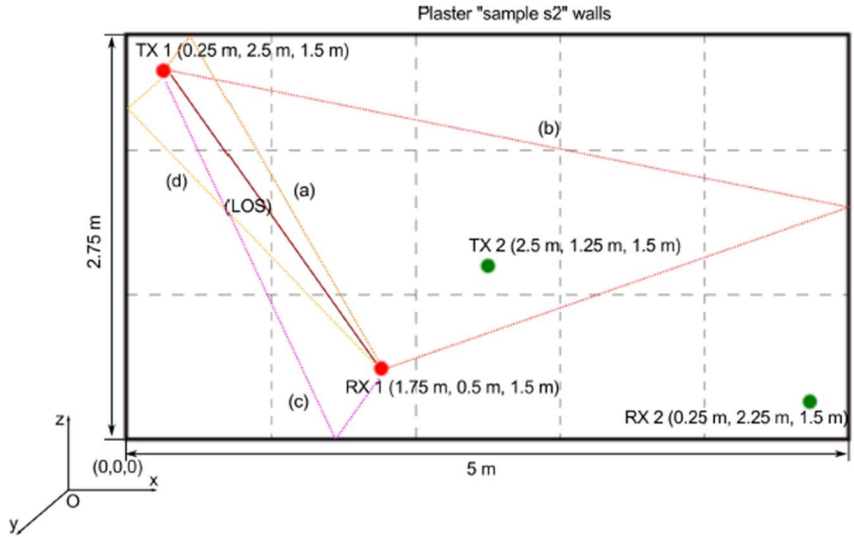

Fig. 4. A deterministic NLOS scenario

This is a perspective of how the channel modelling of Terahertz can be done for shortrange ultrafast communication. Using much advanced optical sources extremely high data rates can be generated.

Using the above stated formulation and channel modelling the achievable data rates are in the order of Gbps, which is much higher than any of the current wireless service.

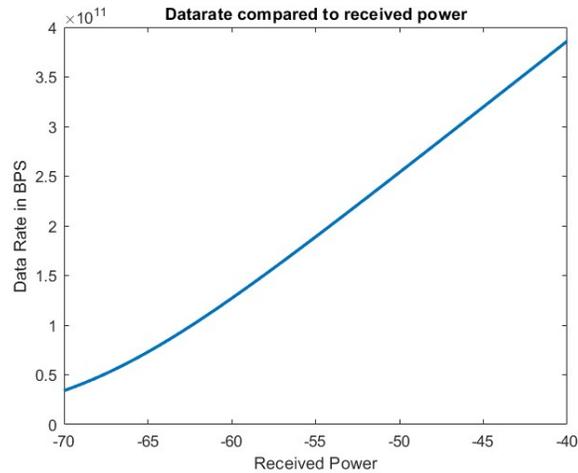

Fig. 5 Shannon Capacity at the given frequency.

We assumed a window of 40GHz which is the system bandwidth, and using Shannon's Channel capacity formula the data rates for SNR of 20dB the following data rates are calculated for a distance of 1 meter.

3. Summary

The increasing demand of bandwidth is pushing industries and researchers to extract ways at higher frequency to satisfy the demands of data handling. This article provides a perspective of achieving much higher data rates for speed intensive data handling. Utilizing the available bandwidth at 300GHz may provide a suitable high speed transmission links for IIoT, payment using credit cards and faster LAN networks.